\documentclass[showpacs,preprintnumbers,amsmath,twocolumn,amssymb]{revtex4}
\usepackage{graphicx}% Include figure files
\usepackage{dcolumn}% Align table columns on decimal point
\usepackage{bm}% bold math

\begin{document}
\title{Modeling for the Dynamics of Human Innovative behaviors}

\author{Ying-Ting Lin$^{1}$}
\author{Xiao-Pu Han$^{2}$}
%\email{xp@hznu.edu.cn}
\author{Bing-Hong Wang$^{1,3,4}$}

\affiliation{$^{1}$Department of Modern Physics, University of Science and Technology of China, Hefei 230026, China\\
$^{2}$Institute of Information Economy and Alibaba Business College, Hangzhou Normal University, Hangzhou, 310036, China\\
$^{3}$College of Physics and Electronic Information Engineering, Wenzhou University, Wenzhou 325035, China\\
$^{4}$The Research Center for Complex System Science, University of Shanghai for Science and Technology, Shanghai 200093, China}
\date{\today}

\begin{abstract}
How to promote the innovative activities is an important problem for modern society. In this paper, combining with the evolutionary games and information spreading, we propose a lattice model to investigate dynamics of human innovative behaviors based on benefit-driven assumption. Simulations show several properties in agreement with peoples' daily cognition on innovative behaviors, such as slow diffusion of innovative behaviors, gathering of innovative strategy on ``innovative centers'', and quasi-localized dynamics. Furthermore, our model also emerges rich non-Poisson properties in the temporal-spacial patterns of the innovative status, including the scaling law in the interval time of innovation releases and the bimodal distributions on the spreading range of innovations, which would be universal in human innovative behaviors.
Our model provide a basic framework on the study of the issue relevant to the evolution of human innovative behaviors and the promotion measurement of innovative activities.
\end{abstract}


\maketitle

\section{Introduction}

The research on Socio-Economic systems rises to be a popular issue and attracts a great amount of researchers in the recent decade \cite{review1,review2,review3,review4}, due to its inherent complicated dynamics and macroscopic phenomenon as the traditional physical system. The foundational method and models of statistical physics, nonlinear physics, complex network, numerical simulation have been applied to the problem of diffusion dynamics, opinion dynamics, traffic flow, cascading failure, evolutionary game, human dynamics and so on. In social development, innovative activity can be regarded as the source of human culture and plays an important role in social advances. To promote the innovative activities of people effectively is always a critical factor for the success of a social system. Here, innovations always refer to new products, goods, services, ideas, online creations, social norms and institutions and so on if they are unknown within a particular group.

After an innovation emerges, it will probably face tough diffusion situation contrasting to the dominative mature technology, owing to kinds of reasons, such as its imperfection or people's stubbornness. The competing course of two different kinds of products or information is a popular topic\cite{strategy, seeding_strategies, information}. Base on a continuous model put forward in 2000, there are a serials of work \cite{techonological_level1, level2_cost, level3, level4, level_cost,level_cost2} discuss when technological development takes place somewhere, how the agents response to neighbors' change and how their reaction contribute to the enhancement of technological level of society.

However, the innovative activities are not only suppressed by difficulties in its spreading, but also in the production of innovation which is the actually the first step of innovative process. Despite extensive discussion on innovations' spread, on the problems of innovative behavior, such as: how the innovative behavior evolves in a interaction system, how the two procedures of innovative activity-innovative behavior and innovation diffusion-affects each other, and most significant, how to promote people's innovative behaviors and the spirit of ingenuity, these relevant discussions still lack in previous studies. The process of innovation bursting has been studied by introducing a technology space based on percolation theory\cite{space1,space2}.

A typical model inspires us greatly is proposed by S. Bornholdt, etc\cite{model} based on the framework with the competition and coexistence of multi-opinions. In this model, innovation is introduced with three basic rules: majority rule when accepting neighbors' opinion, no repeating of old ideas, and tiny chance to raise new idea. The model clearly shows the characteristics of "rapid rise, slow decline, difficulty to replace" of innovation in scientific paradigms' development, and excites our curiosity in the innovative behavior itself.
%Nevertheless, innovation usually is not the only or the best choice of people.
In reality, from the development of discipline and technology, to companies launching new products, or the decision of the topic of a news report, blog, etc., there are always at least two choices for people: to go into a brand new way, or just to improve or follow an exiting one in the system. Thus, there is a problem concerned with multi-choices.

When facing multiple choices, the cost and the payoff from the choice is usually a conclusive consideration. The representative example is evolutionary game, in which one makes different gains according to his own and neighbors' choices, states, or strategies (e.g. to cooperate or to betray), and adopts his next strategy according to the incomes after each play, mostly longing for lower cost and future higher yield. This idea has also been extended to learning process during innovation spread\cite{strategy, techonological_level1,level2_cost, level_cost, level_cost2,PNAS1,technological_level4}. No matter adopting innovative or following behavior, one has to pay cost (in the form of spending time, money, energy, etc. and taking a risk) and can obtain reward (in form of proceeds, reputation, growing interest, etc.). Generally speaking, comparing with the conventionalism and the copycat, innovative behaviors usually face much higher risks and cost but can obtain some extra benefit after the successful innovation, such as the patent income and wide reputations. The following behavior is just the opposite situation with lower risks and lower benefit. Thus the most successful agent (taking highest profits) in a local group may be an innovator or a follower. Hence the innovative and following behavior is just the two choices contradict but interdependent with each other.

In this paper, we propose a modeling framework to study the general feature in the evolution of innovation behavior. Different to most of previous works that mainly discuss the spreading of new technologies or new cultures, we focus on the dynamics in the evolution of innovative behaviors under the assumption that innovators have to pay more price, and that innovative intention is stirred up by huge rewards from the spreading of its innovations. We try to investigate its basic dynamics, and to provide new insights to understand what impacts on peoples' innovative behaviors and how to encourage innovative behaviors effectively.

\section{The model}

\begin{figure}
\center
  \includegraphics[width=7.5cm]{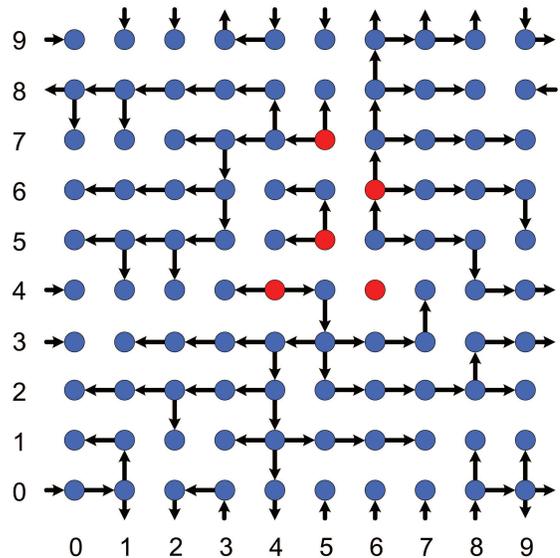}\\
  \caption{(Color online) Illustration of tracks of information spreading on a $10 \times 10$ lattice in a time step $t$. All of the tracks begin with the I agents (red nodes) and respectively strings several F agents (blue nodes). For the agent $i=(x, y)$, no matter F or I, its benefit $M_{i}$ can be defined as the total number of agents whose information comes from him, for example, $M(6,4) = M(2,1) = 1$, $M(5, 7) = 23$, $M(2,5)=5$. }\label{benefits}
\end{figure}

To investigate the basic feature on the evolutionary dynamics of innovative behaviors, our model discards much of detailed and particular factors, such as the differences on innovative ability, the application values of innovations, and so on. For simplicity, the individuals choose binary strategies: to be innovators (I) or to be followers (F). The innovators bring about significant progresses, and followers make smaller improvement by following innovators.
Two basic assumptions are considered in a general way:
(i) Comparing with the activities of just copying or following other's innovations, innovative activities have much more cost.
(ii) From a general perspective, while someone's result is accepted by more individuals, it is considered to be more successful and gains more rewards. And thus we assume,  for an individual, no matter it is the creator of innovation or not, it can obtain some benefit from its spreading. This benefit is proportional to the spreading range of the innovation from the individual.

Based on the above two assumptions, the evolution algorithms of our model can be described as follows:

(i) In $L \times L$ lattices with periodic boundary condition, each node $i$ represents an individual who holds an arbitrary type of information, and adopts binary strategies $s(i,t)$ to update its information: to be a follower (F) or an innovator (I).

(ii) Our model has two timescales: $t$ and $\tau$. Each time step $t$ includes several sub-steps $\tau$.

(iii) At each sub-step $\tau$ during a time step $t$, we select an arbitrary agent $i$ who has not updated on its information at the current $t$. If $i$ is an I agent, it puts forward new information (or say "innovation") that is marked as $m_i(t)$. If node $i$ is a F agent, it randomly contacts with one of his neighbors $j$. If the neighbor's information type is the new one that is created in the current $t$ ($m_j(t,\tau) = m_k(t)$, say, here $k$ is the I agent who puts forward this information type), the F agent $i$ updates its information as the neighbor's ($m_i(t, \tau+1) = m_j(t,\tau) = m_k(t)$).
Noticed that agent, no matter I or F, can only updates its information one time during each $t$, denoting the competition in the spreading of information (innovations).

(iv) For each update on its information, no matter I or F agent it is, the agent consumes some cost. Here we set the cost $C_I =a$ for I agents, and $C_F=1$ for F agents, where $a \geq 1$ means the relative cost of innovative activity and is the main parameter of the model.

(v) When all the agents update their information, we can calculate the payoff of each agent. For agent i, no matter I or F, it can obtain the benefit $M_i$ from the spreading ranges of its information, which is defined as the total number of followers whose information spreads from it, as shown in Fig. 1 and described in Ref. \ref{benefits}.
And thus the final payoff of agent $i$ is the difference of its benefit and cost:
\begin{equation}
    P_i=M_i-C_i=\left\{
    \begin{array}{cc}
M_i-1, &(i \in F), \\
M_i-a,&(i \in I).
    \end{array}
    \right.
\end{equation}

(vi) Agents update their strategies according to their and their neighbors' payoffs. The agent $i$ compares payoffs of its own and neighbors', and then decides its new strategy $s(i,t+1)$ as the same as the winner's. In case there is more than one winner, we randomly choose one of them (node $i'$, say) to learn from. The rule therefore can be written as:
$s(i,t+1)=s(i',t)$, if $P_{i'} \geq P_{k}$($k\in \Omega$, where $\Omega$ is an assembly of $i$ and his four neighbors).
And then the system goes into the next loop ($t = t +1$ and reset $\tau = 0$) and repeats the above rules.ve rules.

\section{Simulation Results}

\begin{figure*}
\center
  \includegraphics[width=7.75cm]{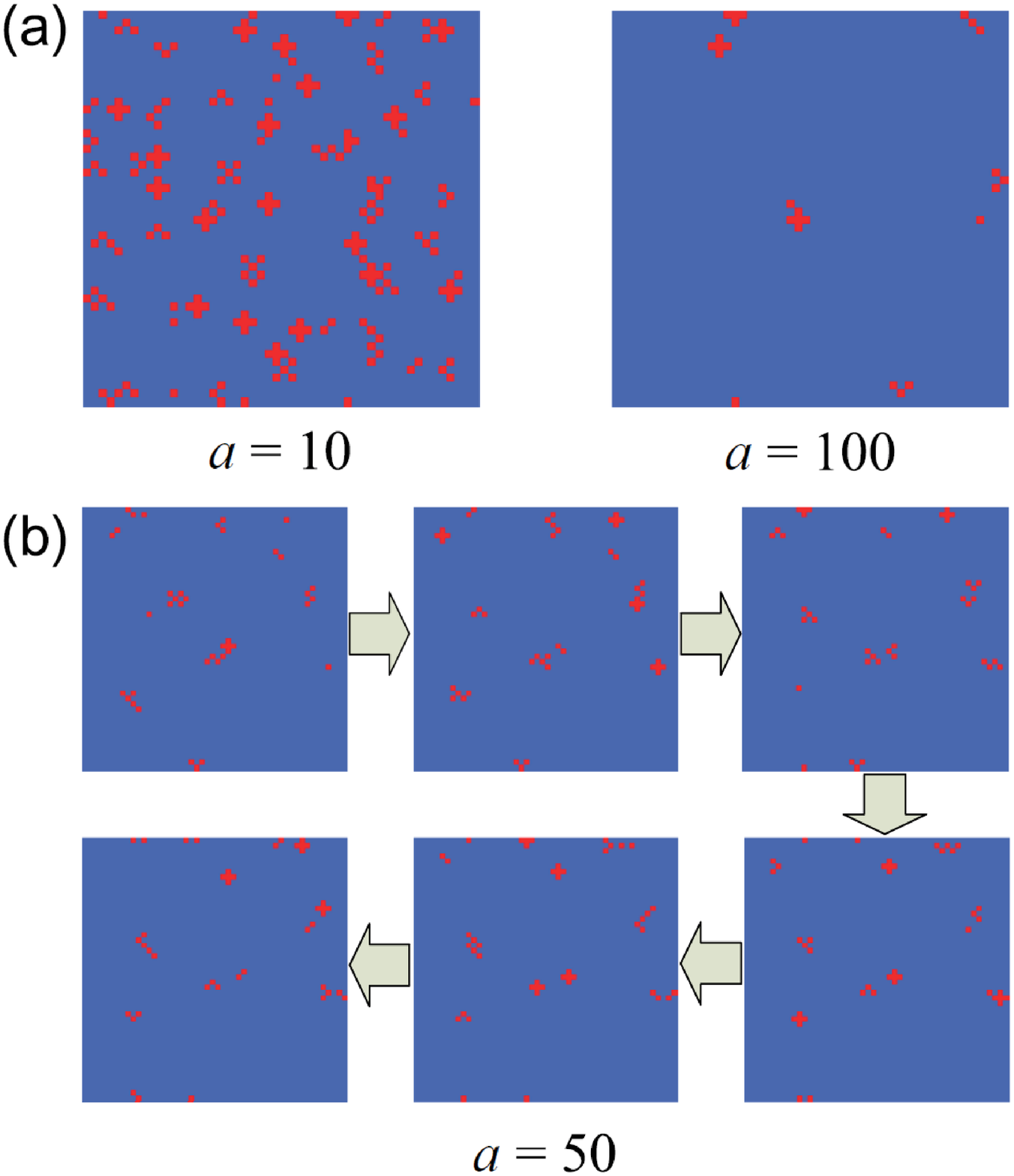}
  \includegraphics[width=5.6cm]{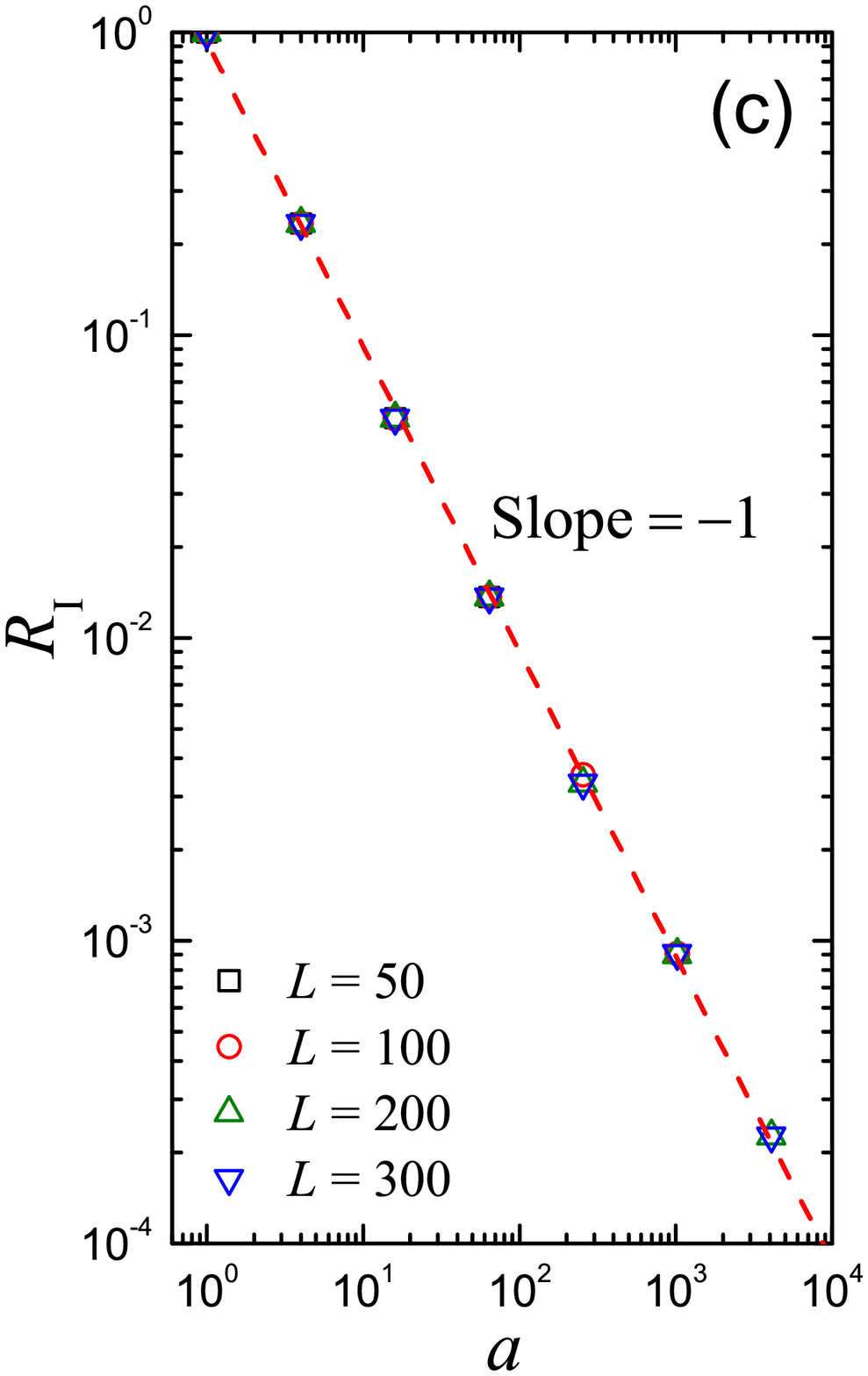}\\
  \caption{(Color online) (a) shows the typical global patterns of in the evolution of the model for different cost value $a$ when $L = 50$. (b) shows the evolution of the pattern in six consecutive time steps for $a = 50$ when $L = 50$. (c) The average proportion of innovators $R_I$ vs. the innovation cost $a$, which obeys a scale-free behavior $R_I(a)\propto a^{-1}$. Results averaged over 50 independent runs, and each run is over 2000 time steps after initialization. }\label{pattern}
\end{figure*}

In our simulations, we set the strategies of nodes randomly as I or F at the first time step $t = 0$. The innovators' number $n_I$ will rapidly achieve an evolutionary stable state in few time steps and then oscillates on a tiny scale.
In the following discussions we set the first 500 time steps (from $t = 0$ to $t = 500$) as the initialization process, and all the statistical properties discussed below are calculated after the initialization process.

The typical patterns in the evolution of the model show that the number of I agents decreases as the increasing of the cost $a$, and these I agents usually gather to several small clusters (Fig. \ref{pattern} (a) and (b)), which have two basic structural elements: the cross plaid and the chessboard-like plaid. The evolution of patterns in several consecutive time steps shows that the two basic structural elements usually alternatively vary and fluctuate in small regions (Fig. \ref{pattern} (b)). This gathering picture is generally in agreement with the reality that most of the sci-tech innovations are generated by some centers (e.g. higher education institutions and big companies). Notice that the present model dose not consider the mechanism that can directly drive the gathering, such as the convenience on the exchange of resource and knowledge. Here the emergence of clusters is the result of the co-effect of the randomly spreading of innovations and social learning on the strategies of local winners.

More deep simulations show that the average proportion of the I agents $R_I$ vs. the cost $a$ of innovation as a prefect inverse proportion function $R_I(a)\propto a^{-1}$ $(a < N)$ which is almost independent to the size of lattices, as shown in Fig. \ref{pattern}(c). This inverse proportion relationship can be analyzed as follows:
According to the algorithms of the model, the critical condition that the I agents stably exists is that the average payoff $\langle P_I\rangle$ of I agents is equal to their nearest neighboring F agents. The average payoff of these nearest neighboring F agents is $\frac{\langle M_I\rangle}{\langle m_F\rangle}-1$, where $\langle m_F\rangle$ is the average number of the nearest neighboring F agents of each I agent, obviously $0\leq m_F\leq4$.
So we have $\langle P_I\rangle=\langle M_I\rangle-a=\frac{\langle M_I\rangle}{\langle m_F\rangle}-1$. We thus can easily understand that $\langle m_F\rangle \geq 2$ and $M_I \geq 2(a-1)$ when $a>1$, namely in the evolutionary stable status, each I agent averagely has at least two spreading branches. Considering the critical condition $\sum M_I \simeq 2n_I(a-1) = N$, we therefore have $R_I = n_I N^{-1} \sim (a-1)^{-1}$, and thus $R_I \sim a^{-1}$ for $a\gg 1$. It shows that the cost of innovation is indeed the major factor on the spreading of innovative behaviors. However, the slow-decreasing inverse proportion relationship implies that innovative behaviors can exist even though the cost is much higher.

\begin{figure}
\center
  \includegraphics[width=8.6cm]{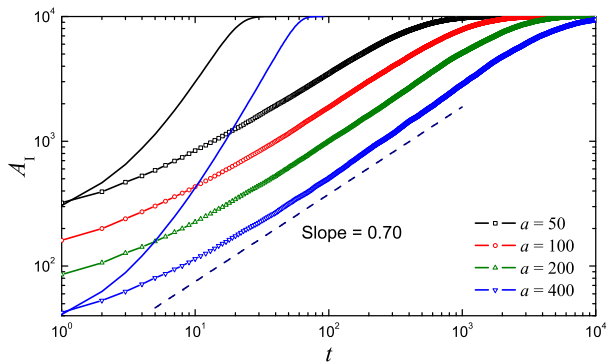}\\
  \caption{(Color online) The growth of the visited area $A_I(t)$ of I strategy for different $a$ (the curves with symbols), comparing with the scenario with random-walk agents (the black and blue curve). $A_I(t)$ also represents for the number of agents who have ever held I strategy before. Parameter settings are $L=100$, and the results are averaged by 20 independent runs. }\label{growth}
\end{figure}

Fig. \ref{pattern}(b) indicates that the clusters of I strategy usually lingers in small regions and is difficult to diffuse out. To give a visible sense, we compare the diffusion speed of the I strategy in our model with that under random-walking scenario. In order to insure the same initial states of two simulations, we make the system begin to evolve with two separated rules after the system stabilized.
In the random-walk scenario, at every step $t$, each agent copies arbitrary one of his neighbors' strategy as his next strategy. In the process, the reappearance of the same strategy is also allowed for a certain agent.
As shown in Fig. \ref{growth}, in our model, this growth of the area $A_I$ that I strategy has visited generally obeys the power form with exponent about $0.7$, which is unsensitive to the value of the cost $a$ and much lower than the random-walk scenario.
This quasi-localized picture indicates that the diffusion of the innovative behavior is quit slow, and the clusters of I agents can keep stable during many time steps. In other words, the small regions that I clusters are lingering can be regarded as ``innovative centers", in which agents have much higher probability to be I strategy.

\begin{figure*}
\center
  \includegraphics[width=15cm]{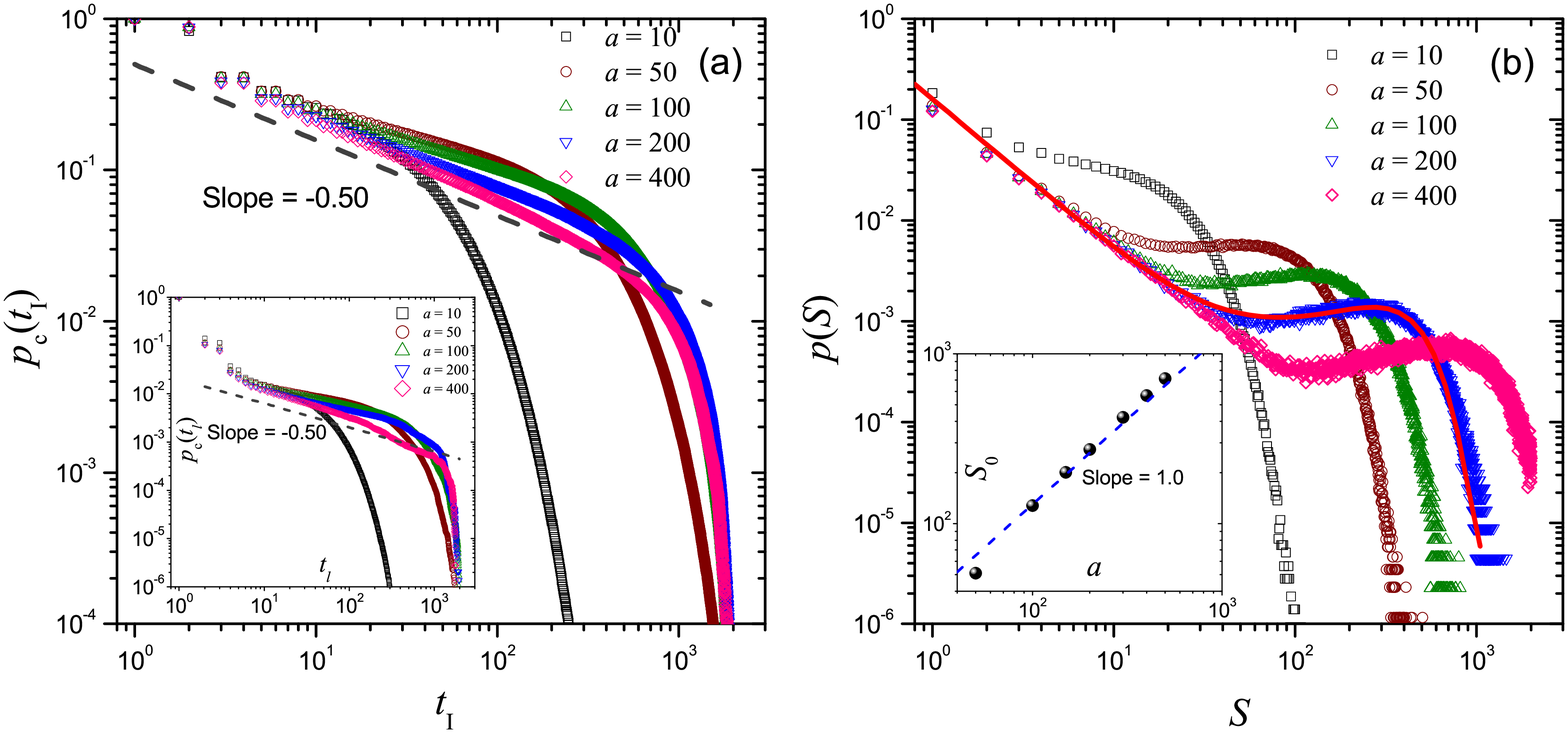}\\
  \caption{(Color online)
  (a) The cumulative distribution $p_c(t_I) $ of time intervals $t_I$ between each two consecutive adoptions of I strategy for different $a$. The inset shows the cumulative distribution $p_c(t_l)$ of the duration $t_l$ that an agent keeps I strategy. The dashed dark gray lines in the main panel and inset indicate the scale-free property with exponent $-0.50$.
  (b) The distributions $p(S)$ of innovators' opinion diffusing scope $S$ for different $a$. The curve of $a=200$ is well fitted by Eq. 2 (The red line), where is the mixed function of power-law and normal distribution with fitting parameters $S_0 = 274.0$, $\gamma = -1.52$, $w = 211.8$, $b = 0.0013$, $c = 0.16$, $d = 2.57\times 10^{-8}$. The inset shows the fitting parameter $S_0$ vs. $a$, where the dashed blue line denotes the proportional relationship $S_0 \sim a$.
  Simulations run for $L=100$ and over 2000 time steps after initialization, and all the results are averaged over 20 independent runs.}\label{scaling}
\end{figure*}

We also investigate the temporal-spacial properties of the information spreading process and surprisingly observe rich abnormal scaling laws and quasi-localized properties, especially when the cost $a$ is higher.

Firstly, the temporal behavior on the change of individual's strategies also shows obvious scaling properties. As shown in Fig. \ref{scaling}(a), the cumulative distributions $p_c(t_I)$ of the interval time steps $t_I$ between two consecutive adoptions of I strategy trend to power-law-like form along the growth of $a$. The exponent of $p_c(t_I)$ is about $0.5$ (corresponds to the exponent $-1.5$ of its probability density distribution), which is a typical exponent on the non-Poisson temporal properties of many real-world human activities \cite{barabasi2005, Vaq2006}. Moreover, the duration $t_l$ that an agent keeps in I strategy obeys much similar scaling property (inset of Fig. \ref{scaling}(a)).
These temporal scaling properties indicate that individual's innovative behaviors likes to burst in short time, and the I strategy usually repeatedly visits a same individual in a short term, associated with few long time durations keeping on F status. Since many online information releases activities somewhat can be regarded as some form of "innovative behavior", $t_I$ also is the time interval of two consecutive information releases, and the scaling interval time distribution is much possible relevant to the universal characters in the temporal patterns of human creative activities due to the similarities on the patterns \cite{Bao, Wang, Song}.

Secondly, the distributions $p(S)$ of the spreading area $S$ of each information in each time step $t$ shows a bimodal type mixed by a power law and a normal distribution:
\begin{equation}
p(S)=b\exp(\frac{-(S-S_0)^2}{2w^2})+cS^{\gamma} +d.
\end{equation}
Since the information is created by I agents, obviously $S_i = M_{I, i}$ for the I agent $i$. As shown in Fig. \ref{scaling}(b), the range of the front scaling section on $p(S)$ grows along the cost $a$, and the exponent $\gamma \simeq -1.5$, which is unsensitive for all the parameters. The I agents in the low I-strategy-density region have more probability to create new information with higher $S$, due to the random spread of information in the system free of competition from other I agents in the same innovative cluster.

A noticeable feature on $p(S)$ is its second peak, which generally corresponds to the peak on the normal distribution section. The insert in Fig. \ref{scaling}(b) shows that the value of the second peak $S_0$ is almost linearly correlated on the cost $a$. Considering $R_I(a)\propto a^{-1}$, we have $S_0 \sim R_I^{-1}$; namely $S_0$ is much correlated to the average spreading area of each information type. This result implies a localized picture on the dynamics in the system: in most of its lifetime, each cluster of I agents drives the evolution in its surrounding field and has almost independent dynamics to other I clusters, unless divides to two clusters, merges to another one, and dies out. In other words, the long-range correlations that drive the emergence of the scaling property is broken and limited in the respective innovative region with I cluster as the center.

\section{Discussion}

The present model discusses the benefit-driven evolution of innovative behaviors combining with the spreading of information and social learning rules, based on the following three basic assumptions: i) the higher cost of innovative behavior; ii) people's innovative behaviors is mainly driven by their benefit; iii) the wider the spreading range of innovation is, the more reward its innovator will obtain. These assumptions have strong daily-life basis and obey people's intuitive feeling on innovative behaviors.

Our model actually has only one effective parameter: the relative cost $a$ of innovative activity. The highlights of the results of our model are the following basic features:

Firstly, the relative cost of innovative activity deeply impacts on the dynamics of innovative behaviors. The effects of the cost are not only shown in the number of nodes with innovative status, but also on the temporal-spatial patterns of individuals' innovative activities and the spreading of innovations (information). To reduce this cost is an efficient way to promote innovative activities.

Secondly, the dynamics of the model shows quasi-localized property, which is exhibited in several phenomena: the gathering and slow diffusion of I status, and the bimodal distributions on spreading ranges of information. This localized picture indicates the emergence of ``innovative centers'', generally matching the reality that most of important innovations are mainly created by few organizations. Moreover, these ``innovative centers'' are emerged naturally in our model due to the absence of the diversity in innovative capability, implying the quasi-localized property would be an intrinsic character in the dynamics of innovative behaviors.

Finally, rich non-Poisson temporal-spatial properties emerges in the situation with large cost (Fig. \ref{scaling}). The scaling law on the interval time of information releases is relevant to the patterns of real-world human creative activities and online information releases \cite{Bao,Song}, implying a novel type of dynamics in the emergence of fat-tailing properties in human behaviors. The power-law section in the distributions of spreading ranges of information also is qualitatively similar to the empirical analysis of the copy and follows of online information.

In summary, we provide a general framework to the modeling studies on the dynamical evolution of human innovative behaviors. This much simplified model obtains rich phenomena matching to the real-world innovative activities, indicating it is indeed effective to show the basic dynamics in the evolution of innovative behaviors and would be helpful in further studies on the promotion of innovative activities.

\begin{acknowledgments}
This work was funded by the National Important Research Project (Grant No. 91024026), the National Natural Science Foundation of China (No. 11205040, 11105024, 10975126, 11275186). the Major Important Project Fund for Anhui University Nature Science Research (Grant No. KJ2011ZD07) and the Specialized Research Fund for the Doctoral Program of Higher Education of China (Grant No. 20093402110032).
\end{acknowledgments}

\end{document}